\documentclass[conference]{IEEEtran}

\usepackage{epsfig,amsmath,amssymb,epsf,amsthm,scalefnt,multirow,subfig}
\usepackage{caption}
\usepackage{xcolor}
\usepackage{float}
\usepackage{cite}
\usepackage{psfrag}
\usepackage{cleveref}

\newtheorem*{lemma*}{Lemma}

\theoremstyle{definition}

 \def\cN{{\mathcal{N}}}

\def\b0{{\pmb{0}}} 

\IEEEoverridecommandlockouts
\IEEEpubid{\makebox[\columnwidth]{978-1-7281-4490-0/20/\$31.00 \copyright~2020~IEEE }
	\hspace{\columnsep}\makebox[\columnwidth]{ }}

\begin{document}

\title{Noncoherent OOK Symbol Detection with Supervised-Learning Approach for BCC}

 \author{\IEEEauthorblockN{Jihoon Cha$ ^\ast $, Junil Choi$ ^\ast $, and David J. Love$ ^\dagger $}
\IEEEauthorblockA{$ ^\ast $School of Electrical Engineering, KAIST, Daejeon, Korea\\
Email: charge@kaist.ac.kr, junil@kaist.ac.kr\\
$ ^\dagger $School of Electrical and Computer Engineering, Purdue University, West Lafayette, IN\\
	Email: djlove@purdue.edu}}

\maketitle

\begin{abstract}
There has been a continuing demand for improving the accuracy and ease of use of medical devices used on or around the human body. Communication is critical to medical applications, and wireless body area networks (WBANs) have the potential to revolutionize diagnosis. Despite its importance, WBAN technology is still in its infancy and requires much research. We consider body channel communication (BCC), which uses the whole body as well as the skin as a medium for communication. BCC is sensitive to the body's natural circulation and movement, which requires a noncoherent model for wireless communication. To accurately handle practical applications for electronic devices working on or inside a human body, we configure a realistic system model for BCC with on-off keying (OOK) modulation. We propose novel detection techniques for OOK symbols and improve the performance by exploiting distributed reception and supervised-learning approaches. Numerical results show that the proposed techniques are valid for noncoherent OOK transmissions for BCC.
\end{abstract}


\section{Introduction}\label{sec1}
There has been significant growth in research and commercialization of medical devices over the last decade, with a particular focus on integrated technologies that increase life expectancy. Wireless communications is critical to many medical devices \cite{IEEE802.15.6,Movassaghi:2014,Latre:2011,Shinagawa:2004}, and the IEEE 802.15.6 task group was organized for the standardization of wireless body area networks (WBANs). WBANs can also be employed in non-medical applications including gaming devices and mobile applications linked to physical activities. To enable increased use, improvements in WBAN power consumption, device compactness, dependence on additional equipment, and other areas are needed \cite{Cotton:2014,Hanson:2009}.

The IEEE 802.15.6 task group defines three physical layers (PHYs) for WBANs: narrowband (NB), ultra-wideband (UWB), and body channel communication (BCC) \cite{Cavallari:2014}. When wireless communication in a WBAN system is conducted on or inside a human body, the communication channel is significantly affected by the body itself. If modeled probabilistically, the channel operating on or near the body will follow a much different distribution than those typically used for wireless channels. Efforts for statistical channel characterization in such a different environment were performed for NB and UWB \cite{Sangodoyin:2018,Ambroziak:2016,Smith:2011}. The theoretical performance analysis and verification of the channel for these PHYs such as channel capacity, power allocation, and outage probability were conducted in \cite{Razavi:2019,Cheffena:2015}.

The system for BCC operates in lower carrier frequencies roughly in 5-50 MHz \cite{Cotton:2014}. Unlike other WBAN communication exploiting NB and UWB, BCC exploits not only the skin but the whole body as a medium for communication. The communication using NB or UWB suffers from blockage of the body resulting in large path loss \cite{Fort:2006}. BCC, on the contrary, utilizes the higher conductivity of the body than that of air, and transceivers for BCC consume lower power by using an electrode \cite{Bae:2012}. BCC is challenging, however, because the channels used for communication vary significantly from standard wireless channel models. One possible avenue for understanding the channel is to use a simple electronic circuit model, which operates in the way of capacitive and galvanic coupling \cite{Seyedi:2013,Callejon:2012,Bae:2012}. Typical stochastic channel modeling for BCC, however, has not been developed because a number of parameters configure the channel conditions, which makes the classification and analysis of the channel difficult.

One common technique to overcome channel fading is the use of diversity. Distributed diversity techniques, which utilize multiple receive nodes distributed over a geographic area, have been shown to be a low-cost and power-efficient solution to achieve performance achievements \cite{Choi:2015,Ibrahim:2017,Brown:2014}. Other than a small number of works (e.g., \cite{Ouyang:2009}), distributed multiple antenna techniques have received little attention for use in or around the body. There is clearly a need to understand how multiple receive nodes could be used in combination with BCC.

In this paper, we consider a realistic communication model for BCC exploiting distributed reception to obtain spatial diversity. On-off keying (OOK) modulation is used to send binary signals, which is a default mode of the WBAN standard \cite{IEEE802.15.6}. Distributed reception is conducted where a fusion center is wired with the other receive nodes and collects the necessary information for symbol detection. Because the body's natural circulation and movement make accurate channel estimation difficult, we assume that a noncoherent model must be employed. We propose three novel techniques to detect OOK symbols for BCC. With an assumption of limited resources for symbol detection at the receive nodes, the proposed techniques are based on supervised learning, which was similarly utilized in \cite{Jeon:2018}. Lastly, we verify their performance through numerical simulations.


\section{System Model}\label{sec2}
\begin{figure}
	\vspace{0.1in}
	\centering
	\includegraphics[width=0.60\columnwidth]{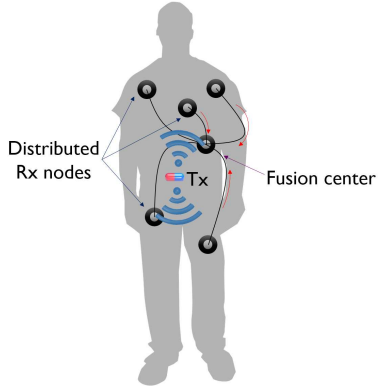}
	\caption{Structure of SIMO model for BCC.}\label{body}
	\vspace{-0.1in}
\end{figure}
We formulate a single-input multiple-output (SIMO) system model for BCC, as depicted in Fig. \ref{body}. The system includes a transmitter and $ K $ distributed receive nodes, all of which have a single antenna. An OOK symbol is transmitted at each time slot by a single-antenna transmitter. The channel between the transmitter and each receive antenna is tightly coupled with the physiology of the human body and the body's movements. For these reasons, we assume that each receive node experiences a different channel model. The chanel gain of each receive node will also quickly fluctuate, which necessitates noncoherent operation.

The received signal at the $ k $-th receive node at time $ n $ is defined as
\begin{align}\label{input_output}
  y_k[n] = \sqrt{P}h_k[n] x[n]+n_k[n],
\end{align}
where $ P $ is the transmit power, and $ h_k[n] $ is the real-valued\footnote{The real-valued channel is widely adopted in optical communication systems \cite{Zhu:2002}. We have extended to the more practical complex-valued channel model in \cite{Cha:2020}.} channel gain of the $ k $-th receive node and is assumed to follow a probability density function $ f_k(h) $. Note that $ h_k[n] $ and $ f_k(h) $ are unknown to both the transmitter and the $ k $-th receive node. The transmitted OOK symbol is denoted by $ x[n]\in\{0,1\} $, and $ n_k[n]\sim \cN\left(0,\frac{N_0B}{2}\right) $ is independent and identically distributed (i.i.d.) noise where $ N_0 $ is the noise spectral density and $ B $ is the bandwidth.

\section{Supervised-Learning-Based OOK Symbol Detection Techniques}\label{sec3}
We assume all the transceivers (i.e., the transmitter, receive nodes, and fusion center) do not have any knowledge of instantaneous or even statistical channel state information (CSI). The property prevents performing typical channel estimation and symbol detection. The proposed detection techniques are, however, still based on a training phase. The fusion center and receive nodes use training signals not to estimate the channel, but to extract useful information in terms of simple statistics such as sample averages to detect the transmitted OOK symbols. We first explain the structure of the training phase and a basic rule of detection, followed by three detection techniques based on supervised learning in detail.

\subsection{Training phase and detection framework}\label{sec3-1}
\begin{figure}
	\vspace{0.1in}
	
	\centering
	\subfloat[][]{\includegraphics[width=0.85\columnwidth]{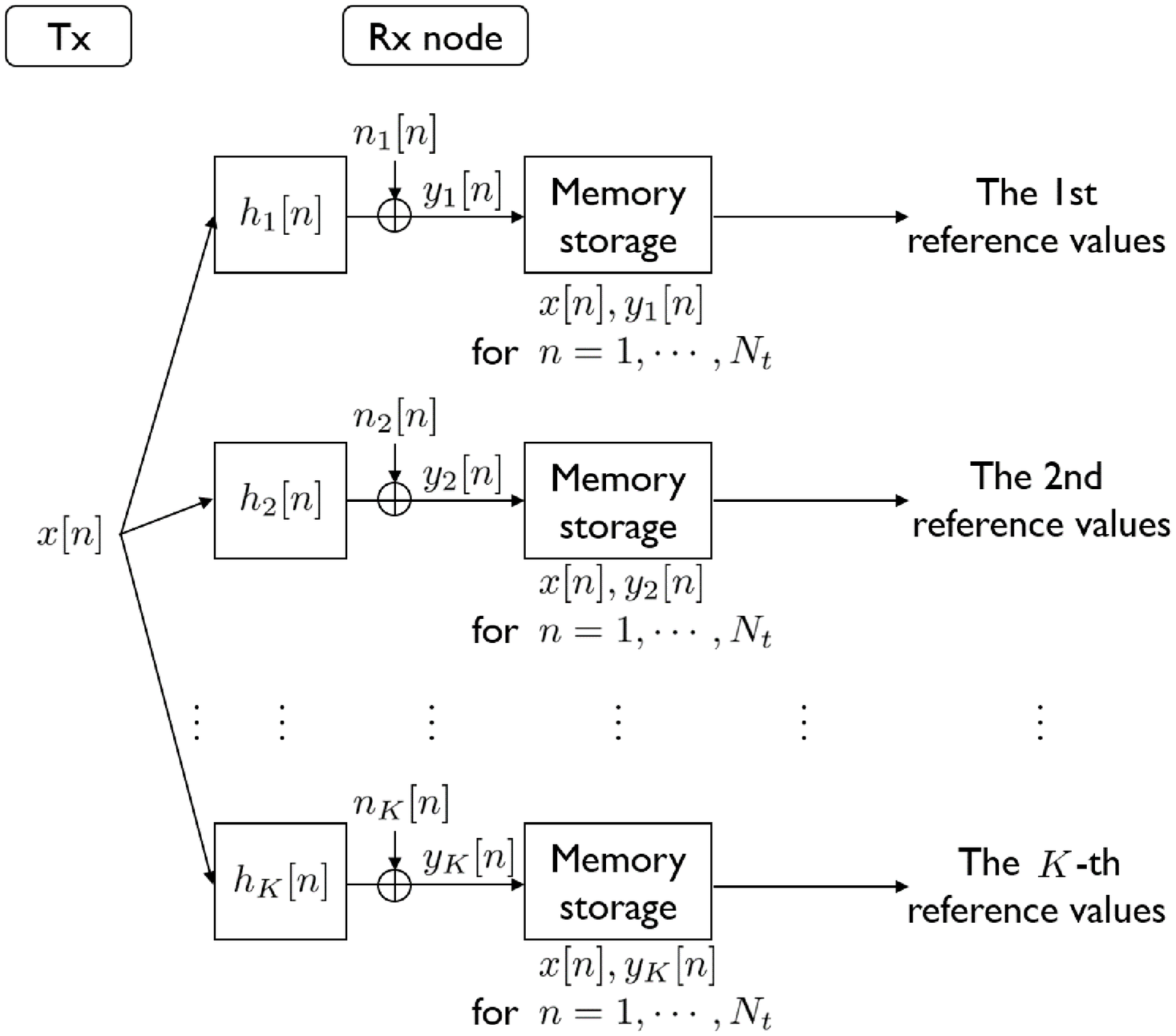}
		\label{training}}
	\hfil
	\subfloat[][]{\includegraphics[width=0.85\columnwidth]{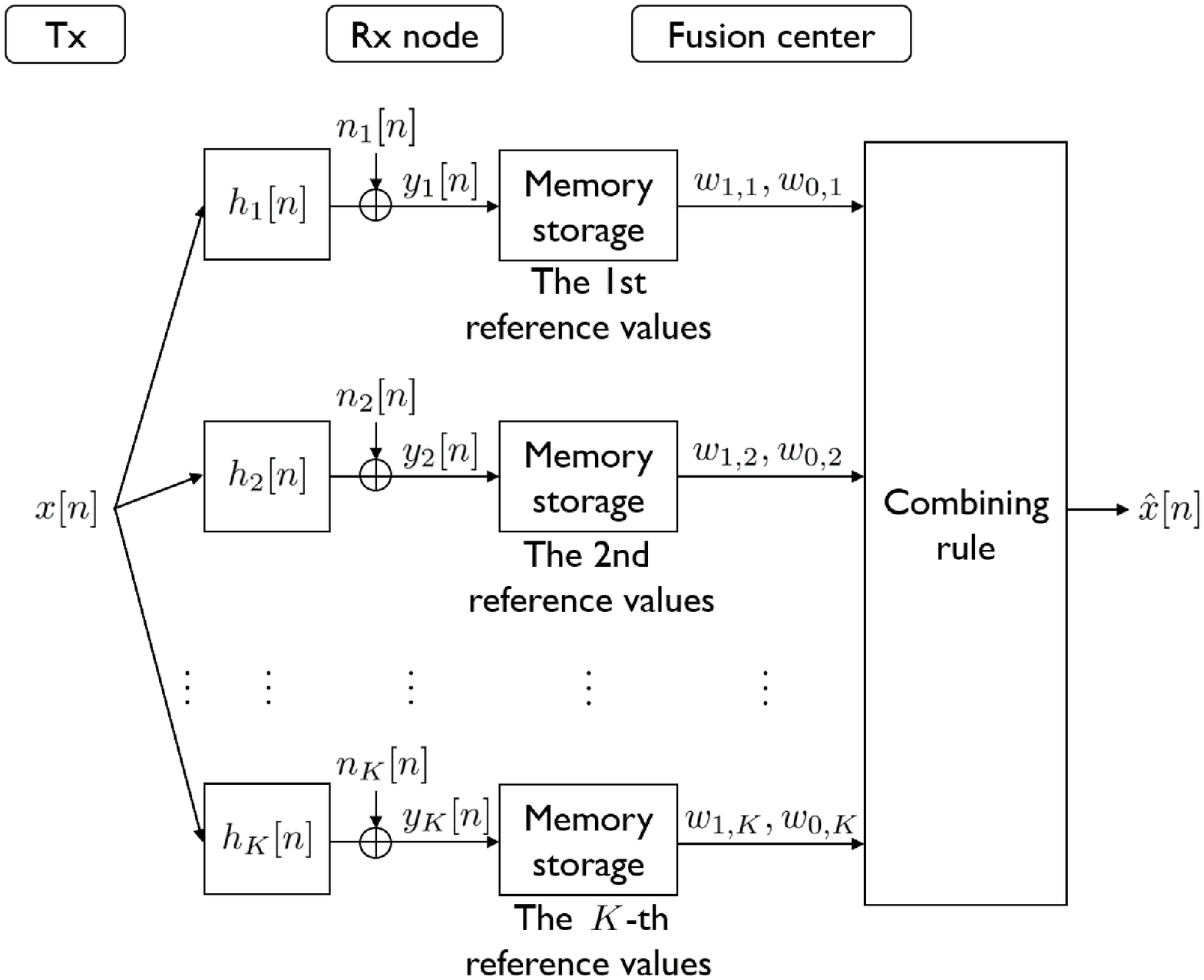}
		\label{data}}
	\caption{The process of OOK symbol detection that consists of \protect\subref{training} training phase with $N_t$ training time slots and \protect\subref{data} data transmission phase.}
	\label{process}
	
	\vspace{-0.1in}
\end{figure}\
The overall process of training and data transmission is depicted in Fig. \ref{process}. In the training phase, the transmitter sends the training symbols known to all receive nodes as
\begin{align}\label{training_signal}
x[n] = \begin{cases}
1, &\text{for } n=1,2,\cdots,\frac{N_t}{2} \\
0, &\text{for }n=\frac{N_t}{2}+1,\cdots,N_t
\end{cases}
\end{align}
during $ N_t $ time slots. Each receive node computes a reference value (depending on a specific detection technique) using the received training signals.

Data transmission is conducted during channel uses $ n = N_t+1 $ and beyond. The fusion center evaluates likelihood with weights that are functions of the reference values determined during the training phase and an instantaneous received data signal. The fusion center detects the OOK symbol as
\begin{align}\label{detection}
\hat{x}[n] = \begin{cases}
1, &\text{for } \sum_{k=1}^{K}{w_{1,k}}[n] > \sum_{k=1}^{K}{w_{0,k}}[n] \\
0, &\text{for } \sum_{k=1}^{K}{w_{1,k}}[n] < \sum_{k=1}^{K}{w_{0,k}}[n].
\end{cases}
\end{align}
In \eqref{detection}, $ w_{1,k}[n] $ and $ w_{0,k}[n] $ are the weights for the $k$-th receive node concerning whether 1 or 0 is transmitted, which are defined by specific detection techniques. The proposed techniques exploit supervised-learning approaches using the reference values to classify the received data signals, which are explained in the rest of this section. It is assumed that each transmitted data symbol is equally likely in this paper.

\subsection{Probability technique}\label{sec3-2}
The probability technique for symbol detection uses empirical conditional probabilities using the received training signals. We first define the threshold amplitude at the $ k $-th receive node as
\begin{align}\label{mean_all}
A_{\mathrm{th},k} = \frac{1}{N_t}\sum_{n=1}^{N_t} \lvert y_k[n]\rvert.
\end{align}
The absolute value of each received training signal at the $k$-th receive node is compared to $A_{\mathrm{th},k}$, which produces a detected training symbol of
\begin{align}\label{detection_threshold}
\hat{x}_k[n]=
\begin{cases} 1, & \text{for }\lvert y_k[n] \rvert \geq A_{\mathrm{th},k}\\
0, & \text{for }\lvert y_k[n] \rvert<A_{\mathrm{th},k}\end{cases}
\end{align}
for $ n=1,\dotsc,N_t $. Using $ \hat{x}_k[n] $, two empirical conditional probabilities are computed for each $ k $. One is for the event of $ \hat{x}_k[n]=1 $ conditioned on $ x[n]=1 $, and the other is similarly for $ \hat{x}_k[n]=0 $ given $ x[n]=0 $. They are written as
\begin{align}\label{cond_prob1}
P_{(1|1),k}=\min{\left(\frac{\sum_{n=1}^{N_t/2}\delta_{\hat{x}_k[n],x[n]}}{N_t/2}, 1-\frac{2}{N_t}\right)}
\end{align}
and
\begin{align}\label{cond_prob0}
P_{(0|0),k}=\min{\left(\frac{\sum_{n=N_t/2+1}^{N_t}\delta_{\hat{x}_k[n],x[n]}}{N_t/2}, 1-\frac{2}{N_t}\right)},
\end{align}
where $ \delta_{\hat{x}_k[n],x[n]} $ indicates Kronecker delta, defined as
\begin{align}\label{kro_delta}
\delta_{\hat{x}_k[n],x[n]}=
\begin{cases}
1, & \text{for }\hat{x}_k[n]=x[n]\\
0, & \text{for }\hat{x}_k[n]\neq x[n].
\end{cases}
\end{align}
These empirical probabilities have discretized values by using the number of correctly detected training symbols. Note that with finite $ N_t $, the empirical probability goes to one as the transmit power increases, which will make the empirical probability useless at high transmit power. To prevent this, the probabilities in \eqref{cond_prob1} and \eqref{cond_prob0} are set to have an upper bound of $ 1-\frac{2}{N_t} $.

In the data transmission phase, a random symbol $ x[n] $ is transmitted, and the amplitude of an  instantaneous received signal is measured at each receive node. Data symbol detection for each $ k $ is conducted with \eqref{detection_threshold}, followed by allocating two weights using \eqref{cond_prob1} and \eqref{cond_prob0} as
\begin{align}\label{weight_prob1}
w_{1,k}^p[n] = \begin{cases}
\log{P_{(1|1),k}}, &\text{for } \hat{x}_k[n]=1 \\
\log{(1-P_{(1|1),k})}, &\text{for } \hat{x}_k[n]=0
\end{cases}
\end{align}
and
\begin{align}\label{weight_prob0}
w_{0,k}^p[n] = \begin{cases}
\log{(1-P_{(0|0),k})}, &\text{for } \hat{x}_k[n]=1 \\
\log{P_{(0|0),k}}, &\text{for } \hat{x}_k[n]=0
\end{cases}
\end{align}
for $ k=1,\dotsc,K $. By combining the weights of \eqref{weight_prob1} and \eqref{weight_prob0} as in \eqref{detection}, the final detected symbol $ \hat{x}^p[n] $ is determined at the fusion center. This technique is similar to a typical likelihood ratio test (LRT) in a binary communication channel \cite{Cover:2006}, but uses empirical probabilities.

\subsection{Deviation technique}\label{sec3-3}
The deviation technique uses the difference value between the amplitude of an instantaneous received signal and the reference values computed during the training phase. Two sample averages are computed with the received training signals for the cases of $ x[n]=1 $ and $ x[n]=0 $, defined as
\begin{align}\label{mean1}
A_{1,k} = \frac{2}{N_t}\sum_{n=1}^{N_t/2} \lvert y_k[n]\rvert
\end{align}
and
\begin{align}\label{mean0}
A_{0,k} = \frac{2}{N_t}\sum_{n=N_t/2+1}^{N_t} \lvert y_k[n]\rvert
\end{align}
for $ k=1,2,\dotsc,K $, which serve as the reference values for the deviation technique.

The weights used in the deviation technique are a function of the received data signal, written as 
\begin{align}\label{weight_dev1}
w_{1,k}^d[n] = \lvert y_k[n] \rvert - A_{1,k}
\end{align}
and
\begin{align}\label{weight_dev0}
w_{0,k}^d[n] = A_{0,k} - \lvert y_k[n] \rvert
\end{align}
for each $ k $. Increasing $ \lvert y_k[n] \rvert $ makes $ w_{1,k}^d[n] $ larger, which results in smaller $ w_{0,k}^d[n] $, and vice versa. The final symbol detection is conducted by computing \eqref{detection} using \eqref{weight_dev1} and \eqref{weight_dev0} to derive $ \hat{x}^d[n] $.

\textbf{Remark 1:} Using the probability and deviation techniques with $ K = 1 $, i.e., having only a single receive node, the fusion center produces the same detection result, since the detecting criterion is simplified to whether or not $ \lvert y_k[n] \rvert $ is larger than $ A_{\mathrm{th},k} $ as in \eqref{detection_threshold} for both techniques. With multiple receive nodes, however, the two techniques can result in different symbol detection results. Considering high transmit power, both $ P_{(1|1),k} $ and $ P_{(0|0),k} $ in the probability technique approach to $ 1-2/N_t $ as in \eqref{cond_prob1} and \eqref{cond_prob0}. One of the two weights in \eqref{weight_prob1} and \eqref{weight_prob0} for all the receive nodes approaches to 0, which is the maximum value of the weights when $ N_t $ is sufficiently large. Once the instantaneous data signal is detected at the receive nodes as in \eqref{detection_threshold}, symbol detection just follows the majority rule at the fusion center, regardless of the value of $ \lvert y_k[n] \rvert $.
The deviation technique, on the contrary, does not just perform the majority rule even with high transmit power, since the weights in \eqref{weight_dev1} and \eqref{weight_dev0} depend on the amplitude of the instantaneous received signal $ \lvert y_k[n] \rvert $.

\subsection{Combination technique}\label{sec3-4}
The combination technique exploits the reference values that have been developed in the previous subsections to compute weights, making a robust detector coping with various channel conditions. The empirical conditional probabilities and sample averages are computed as in \eqref{mean_all}, \eqref{cond_prob1}, \eqref{cond_prob0}, \eqref{mean1}, and \eqref{mean0} during the training phase. Using the received data signal and the reference values with \eqref{mean_all} and \eqref{weight_prob1} to \eqref{weight_dev0}, the weights are defined as
\begin{align}\label{weight_com1}
w_{1,k}^c[n] = 
-\frac{\lvert w_{1,k}^d[n]\rvert ^2}{A_{1,k}}+\frac{\lvert w_{1,k}^d[n]\rvert^2}{A_{\mathrm{th},k}}w_{1,k}^p[n]
\end{align}
and
\begin{align}\label{weight_com0}
w_{0,k}^c[n] = 
-\frac{\lvert w_{0,k}^d[n] \rvert ^2}{A_{0,k}}+\frac{\lvert w_{0,k}^d[n] \rvert ^2}{A_{\mathrm{th},k}}w_{0,k}^p[n]
\end{align}
for each $ k $.

Focusing on \eqref{weight_com1}, the first term of $ w_{1,k}^c[n] $ is a scaled and squared version of the weights from the deviation technique. Using the scaled form of the first term is appropriate to the asymmetric magnitude distribution of the received data signal for each case of $ x[n] = 1 $ and $ x[n] = 0 $. The second term of $ w_{1,k}^c[n] $ considers the probability technique. Since $ w_{1,k}^p[n] $ is defined as the logarithm of empirical probabilities in (9), the premultiplied value to $ w_{1,k}^p[n] $ serves as an exponent for the probabilities as a base, which adjusts degree of penalty. Due to the premultiplied value for the second term, the contribution of the two terms to the weights is balanced. The fusion center detects the final symbol $ \hat{x}^c[n] $, combining \eqref{weight_com1} and \eqref{weight_com0} as in \eqref{detection}.

\section{Numerical Results}\label{simul}
\begin{figure}
	\centering
	\includegraphics[width=0.91\columnwidth]{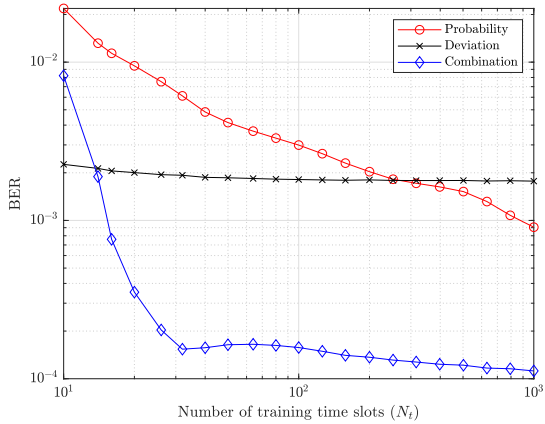}
	\caption{BERs for the group of weak channels with $ K = 6 $ by changing the number of training time slots from 10 to 1000 for fixed transmit power of 10 dBm.}\label{nt101000}
	\vspace{-0.1in}
\end{figure}

\begin{table}[!t]
	
	\vspace{0.1in}
	\renewcommand{\arraystretch}{1.2}
	\captionsetup{labelsep = newline,justification=centering,font={footnotesize,sc}}
	\caption{Considered channel probability distributions}
	\label{distribution}
	\centering
	\begin{tabular}{c c c}
		\hline\hline
		$ f_k(h) $ & Distribution model & Condition\\
		\hline
		$ f_1(h) $ & Burr $ ([4.71*10^{-7},2.43,5.61]) $ & weak\\
		$ f_2(h) $ & Burr $ ([9.32*10^{-7},3.88*10^1,5.52*10^{-1}]) $ & strong\\
		$ f_3(h) $ & Burr $ ([2.29*10^{-8},1.21*10^1,5.07*10^{-1}]) $ & weak\\
		$ f_4(h) $ & Burr $ ([5.63*10^{-6},2.40*10^1,3.97*10^{-1}]) $ & strong\\
		$ f_5(h) $ & Weibull $ ([1.76*10^{-6},3.88]) $ & weak\\
		$ f_6(h) $ & Burr $ ([3.83*10^{-7},7.06,1.26]) $ & weak\\
		$ f_7(h) $ & Burr $ ([1.31*10^{-6},5.25,1.47]) $ & weak\\
		$ f_8(h) $ & Weibull $ ([1.01*10^{-6},4.05]) $ & weak\\
		$ f_9(h) $ & Burr $ ([7.76*10^{-6},9.71,7.87]) $ & strong\\
		\hline
	\end{tabular}

\vspace{-0.1in}
\end{table}
We evaluate the performance of the proposed detection techniques by computing uncoded bit error rate (BER) with Monte-Carlo simulations. Presuming that channels in BCC have inconsistent probabilistic models, we make use of the channel realization following the probability distributions in Table \ref{distribution} that are extracted from \cite{Razavi:2019}. The noise spectral density $ N_0 $ and the bandwidth $ B $ are set to -174 dBm/Hz and 100 kHz corresponding to that of the distribution models in \cite{Razavi:2019}. Classification of the distribution models as ``strong'' and ``weak'' in Table \ref{distribution}, which depends on the channel condition, will be discussed later. We consider a limited situation where the channel distribution between the transmitter and each receive node does not change for the time of interest. However, both the instantaneous channel value, which changes in every time slot, and the channel distribution are not known to all the transceivers, which makes the system noncoherent. The number of training time slots is set to $ N_t = 50 $ for all simulations except the last one in Fig. \ref{nt101000}.

\begin{figure}
	\centering
	\includegraphics[width=0.91\columnwidth]{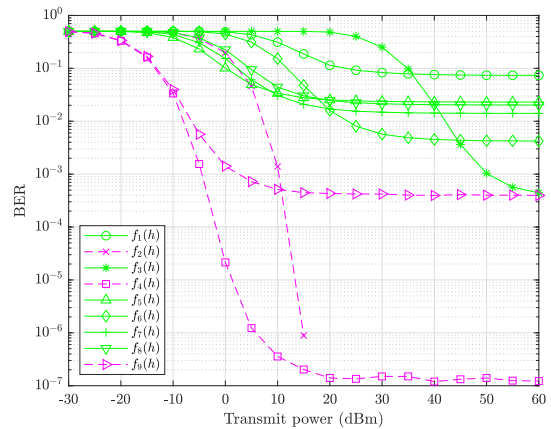}
	\caption{BERs using the probability technique with a single receive node for nine different channel distribution models.}\label{k1_123456789}
	
	\vspace{-0.1in}
\end{figure}
Depending on channel condition, the probability technique with a single receive node shows all different BER performance. The result is shown in Fig. \ref{k1_123456789} where we exploit nine channel distribution models separately. The probability technique shows a bounded tendency of BERs with high transmit power. As transmit power increases, the reference values used in the technique are given less effect from the noise. The reference values, however, would remain fixed after a certain transmit power level as discussed in \textbf{Remark 1} in Section \ref{sec3-3}, which do not handle fast-varying channels. Meanwhile, we purposely classify the channel distributions into the group of strong channels or the group of weak channels. The solid green lines in Fig.
\ref{k1_123456789} correspond to the weak channels, and the dashed pink lines represent the strong channels. The BERs of the strong channels start to decrease from small transmit power, achieving lower bound with high transmit power, and vice versa. Some channel distributions including the group of strong channels have large mean or small variance, which leads to reliable communication through the channels with small randomness, and vice versa.

\begin{figure}
	\centering
	\includegraphics[width=0.91\columnwidth]{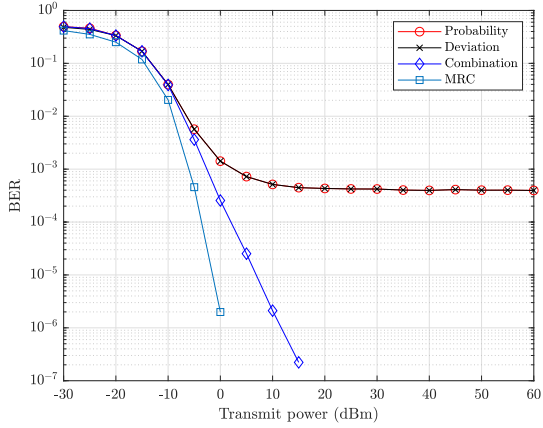}
	\caption{BERs using the three detection techniques with $ K = 1 $ where $ f_9(h) $ is used for coherent and noncoherent cases.}\label{k1_9}
	\vspace{-0.1in}
\end{figure}
In Fig. \ref{k1_9}, we focus on one of the channels $ f_9(h) $ to clearly verify the disadvantage of noncoherent detection with the performance of the three detection techniques proposed in Section \ref{sec3}. A coherent combiner is used for comparison, operating on the same WBAN deployment scenario with the same channel models. For the coherent case, the fusion center performs maximum ratio combining (MRC) using perfect CSI. The MRC technique shows good performance comparing to all proposed noncoherent detection techniques. The probability and deviation techniques give the same performance as mentioned in \textbf{Remark 1} in Section \ref{sec3-3}. The combination technique resolves the problem of bounded BERs and makes the best performance among the proposed techniques with high transmit power.

\begin{figure}
	\centering
	\includegraphics[width=0.91\columnwidth]{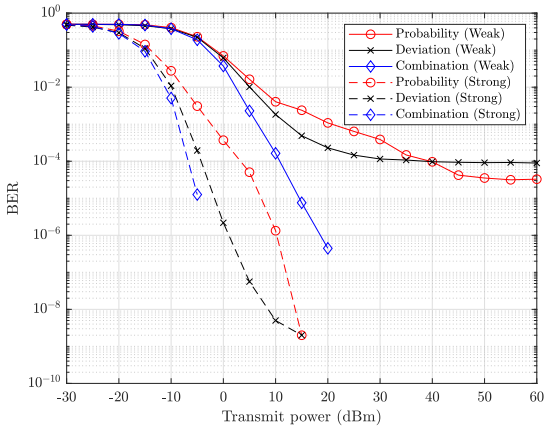}
	\caption{BERs using the three detection techniques for weak-channel-only ($ K=6 $) and strong-channel-only ($ K=3 $) scenarios with solid lines and dashed lines respectively.}\label{k36}
	\vspace{-0.1in}
\end{figure}
The dashed lines in Fig. \ref{k36} show the performance using the group of three strong channels. This favorable situation when only strong channels exist might rarely occur in practice though. Meanwhile, for the solid lines in Fig. \ref{k36}, the fusion center uses six receive nodes following the distributions in the group of weak channels. This channel situation is more conservative than the other. The probability technique has unnatural decrease by combining several conditional probabilities $ P_{(1|1),k} $ and $ P_{(0|0),k} $ with various channels, i.e., as transmit power increases, the empirical probabilities of the weak channels become unreliable as shown from the solid lines in Fig. \ref{k1_123456789}. These unreliable empirical probabilities corrupt likelihoods when the weights for all receive nodes are combined, leading to a rough decrease of BER in some range of transmit power. The deviation technique achieves lower BER than that of the probability technique with low transmit power. The weights that are a function of continuous values are more precise than that using the discretized values, namely, (empirical) probabilities. On the contrary, the deviation technique is inferior to the probability technique with high transmit power. This is because the strong channel in each group would have large variance, which makes the weights using the amplitude of an instantaneous received signal unreliable. Both the probability and deviation techniques exhibit similar performance. The combination technique, on the contrary, outperforms the two techniques in both scenarios.

\begin{figure}
	\centering
	\includegraphics[width=0.91\columnwidth]{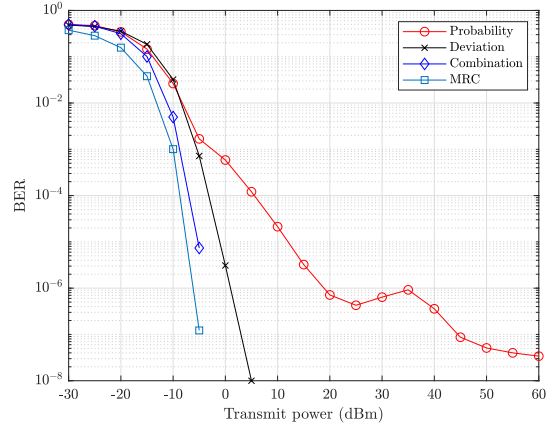}
	\caption{BERs under the various channel distribution models with $ K = 9 $.}\label{k9_123456789}
	\vspace{-0.1in}
\end{figure}
Fig. \ref{k9_123456789} shows the case of using both groups. Compared to the previous two extreme cases with either weak or strong channels only, this is definitely more realistic environment that has a variety of channel condition for each receive node. The probability and deviation techniques for $ K = 9 $ show some performance degradation in some range of transmit power, comparing to the group of strong channels in Fig. \ref{k36}. The combination technique gives no such degradation of BER performance, meeting robustness to the channel conditions, which is comparable to the MRC technique for the coherent channel.

In Fig. \ref{nt101000}, we verify BER performance with changing the number of training time slots $ N_t $ from 10 to 1000 and fixed transmit power of 10 dBm for the group of weak channels. The deviation technique has negligible gain with larger $ N_t $ because the two sample averages in \eqref{mean1} and \eqref{mean0} already converge to certain values even with small $ N_t $. The probability technique, however, has steadily decreasing BER due to the increasing accuracy of the empirical conditional probabilities. The combination technique gives the best performance among the three techniques. Although the transceivers operate with a small value of $ N_t $, the combination technique guarantees adequate BER performance even for weak channels. 

\section{Conclusion}\label{conc}
In this paper, we formulated a realistic BCC system model where OOK symbols are transmitted through fast-varying channels. We showed that it is possible to benefit from multiple distributed receive nodes by exploiting a supervised-learning approach without conventional channel estimation. In this set-up, only a small number of training symbols are used to compute weights that are predefined. The weights are combined at the fusion center for symbol detection.

We proposed three detection techniques called probability, deviation, and combination techniques according to how the weights are defined. Using distributed reception across the multiple receive nodes, it is possible to achieve robust transmissions for the noncoherent BCC system.
The combination technique especially shows good performance by using well-defined supervised-learning approaches with small training overhead.
By designing more effective reference values and weights, it may be possible to make a detector more robust to dynamic fluctuation of channel. A more realistic situation such as multitap channel would be considered as an interesting future work.

\section*{Acknowledgment}
This work was supported in part by the National Research Foundation (NRF) grant funded by the Korean Government (MSIT) under Grant 2018R1A4A1025679 and by the National Science Foundation (NSF) under grants CCF1816013.

\bibliographystyle{IEEEtran}
\bibliography{refs_all}

\end{document}